\documentclass[aip,apl,preprint,showpacs,amsmath,amssymb,floatfix]{revtex4-1}

\usepackage{bm}
\usepackage{graphicx}
\usepackage{siunitx}
\renewcommand{\deg}{$^\circ$}

\begin{document}

\title{Micromagnetic simulations of magnetoelastic spin wave excitation in scaled magnetic waveguides}

\author{Rutger Duflou}
\affiliation{Imec, B-3001 Leuven, Belgium}
\affiliation{KU Leuven, Faculteit Ingenieurswetenschappen, B-3001 Leuven, Belgium}

\author{Florin Ciubotaru}
\email{Florin.Ciubotaru@imec.be}

\author{Adrien Vaysset}
\affiliation{Imec, B-3001 Leuven, Belgium}

\author{Marc Heyns}
\affiliation{Imec, B-3001 Leuven, Belgium}
\affiliation{KU Leuven, Faculteit Ingenieurswetenschappen, B-3001 Leuven, Belgium}

\author{Bart Sor\'ee}
\affiliation{Imec, B-3001 Leuven, Belgium}
\affiliation{KU Leuven, Faculteit Ingenieurswetenschappen, B-3001 Leuven, Belgium}
\affiliation{Universiteit Antwerpen, Departement Fysica, B-2000 Antwerpen, Belgium}

\author{Iuliana P. Radu}
\author{Christoph Adelmann}
\email{Christoph.Adelmann@imec.be}
\affiliation{Imec, B-3001 Leuven, Belgium}

\begin{abstract}
We study the excitation of spin waves in scaled magnetic waveguides using the magnetoelastic effect. In uniformly magnetized systems, normal strains parallel or perpendicular to the magnetization direction do not lead to spin wave excitation since the magnetoelastic torque is zero. Using micromagnetic simulations, we show that the nonuniformity of the magnetization in submicron waveguides due to the effect of the demagnetizing field leads to the excitation of spin waves for oscillating normal strains both parallel and perpendicular to the magnetization. The excitation by biaxial normal in-plane strain was found to be much more efficient than by uniaxial normal out-of-plane strain. For narrow waveguides with widths of 200\,nm, the excitation efficiency of biaxial normal in-plane strain was comparable to that of shear strain. 
\end{abstract}
\maketitle

The control and manipulation of ferromagnetic nanostructures using the magnetoelectric effect has recently received increasing interest.\cite{Fiebig,Eerenstein, Srinivasan, Vaz, Fusil} Of special interest has been the magnetoelectric generation of spin waves due to a potentially much higher energy efficiency than current-based excitation schemes, \textit{e.g.} based on inductive coupling to microwaves or spin torque oscillators. Such magnetoelectric spin wave transducers can therefore be considered as key elements of future low-power magnonic devices.\cite{Khitun1,Radu_MAJ_gates, Dutta_SR, Khitun2} 
 
Magnetoelectric excitation and control of ferromagnetic resonance (FMR) as well as propagating spin waves have been studied in both multiferroic materials, such as BiFeO$_3$, \cite{Rovillain} and magnetoelectric compounds.\cite{WeilerFMR, WeilerSPinPumping, Cherepov, Yu1, Liu1, Liu2, Nan} Magnetoelectric compounds consist of piezoelectric and magnetostrictive layers coupled via strain. The strain generated by the piezoelectric layer upon application of an electric field leads to an effective magnetic anisotropy field in the magnetostrictive layer. The resulting magnetization dynamics can then be described by the Landau-Lifshitz-Gilbert equation. While the physics of magnetoelastic coupling has been established decades ago,\cite{Kittel2, Eshbach} the magnetization dynamics in scaled micromagnetic systems have only recently received attention\cite{Barra} due to their potential for nanoscale magnonic devices. In this paper, we study the generation of spin waves in scaled ferromagnetic waveguides by a local magnetoelastic transducer for different strain geometries. We show that the demagnetizing fields plays a key role and can be used to design efficient magnetoelastic (and thus magnetoelectric) spin wave transducers.

The magnetoelastic energy per unit volume as a function of magnetization $\bm{m} = \bm{M}/M_S$ ($M_S$ being the saturation magnetization of the ferromagnet) and strain tensor $\bm{\varepsilon}$ is given in a first-order approximation by \cite{Kittel}

\begin{equation}
\begin{aligned}
E_\mathrm{mel} = &B_1 \left(\varepsilon_{xx} (m_x^2 - 1/3) + \varepsilon_{yy} (m_y^2 - 1/3) + \varepsilon_{zz} (m_z^2 - 1/3)\right)\\
&+ B_2 (\varepsilon_{xy} m_x m_y + \varepsilon_{yz} m_y m_z + \varepsilon_{xz} m_x m_z).
\label{eq:magnetoelasticenergy}
\end{aligned}
\end{equation}

\noindent Here, $B_1$ and $B_2$ denote the magnetoelastic coupling constants. The corresponding effective magnetic field, $\bm{H} = -\nabla_{\bm{M}} E_\mathrm{mel}/\mu_{0}$, is then given by

\begin{equation}
	\bm{H} = -\frac{1}{\mu_0 M_S}
	\begin{pmatrix}
	2 B_1 \varepsilon_{xx} m_x + B_2 (\varepsilon_{xy} m_y + \varepsilon_{xz} m_z)\\
	2 B_1 \varepsilon_{yy} m_y + B_2 (\varepsilon_{xy} m_x + \varepsilon_{yz} m_z)\\
	2 B_1 \varepsilon_{zz} m_z + B_2 (\varepsilon_{xz} m_x + \varepsilon_{yz} m_y)\\
	\end{pmatrix},
	\label{eq:magnetostrictiveeffectivefield}
\end{equation}

\noindent with $\mu_0$ the vacuum permeability. It is easy to see that for normal strains parallel or perpendicular to the magnetization, the resulting effective field is parallel to the magnetization or zero, respectively. Therefore, no torque $\bm{\tau} \propto \bm{m} \times \bm{H}$ is exerted on the magnetization in such geometries [see Eq.~(S1) in the supplementary material]. 

Several remedies for this issue are possible, such as the application of strains with a shear component, as in the case \textit{e.g.} of Rayleigh surface acoustic waves. \cite{WeilerFMR, Cherepov} Alternatively, a slanted magnetization with respect to a normal strain, either due to an oblique external magnetic field or a canted magnetic anisotropy,\cite{Khitun1} can also lead to nonzero torques. Future magnonic devices may however employ narrow waveguides with dimensions in the 100\,nm range.\cite{Zografos} In such waveguides, when magnetized transversally (along the $y$-direction), the magnetization is nonuniform due to the nonuniformity of the demagnetizing field in this configuration. This is illustrated in Figs.~S1(a) and (b) in the supplementary information for 10\,nm thick waveguides with widths of 200\,nm and 500\,nm, respectively. These---and all following---simulations were performed using the Object Oriented MicroMagnetic Framework (OOMMF).\cite{OOMMF} The parameters of the magnetic waveguide material corresponded to permalloy with an exchange coefficient of $A = 1.3 \times 10^{-11}$\,J/m and a saturation magnetization $M_S = 8\times 10^5$\,A/m.\cite{Yahagi} As in the spin wave excitation studies below, an external transverse magnetic bias field of 50\,mT was applied. The simulation results show both nonzero $x$ and $z$ components near the edges of the waveguides. Nonuniformities were larger for the narrower waveguide, where nonzero $x$ and $z$ components extended into the center of the waveguide.

The magnetoelastic excitation of spin waves in the narrow waveguides was then simulated using the YY\_MEL module within OOMMF.\cite{Yahagi} The external transverse magnetic bias field resulted in Damon-Eshbach-like spin wave modes. The damping constant was assumed to be $\alpha = 0.005$, with a gradual increase to a value of 0.8 within 1\,$\mu$m of the ends of the 10\,$\mu$m long waveguide to avoid backreflection of the spin waves. The magnetoelastic coupling constants were $B_1 = B_2 = 7.85 \times 10^{6}$ J/m$^3$. External strains were applied in a $200\times 200$\,nm$^2$ region in the center of the waveguides with a sinusoidal amplitude modulation with the frequency $f = 8$\,GHz, well above the FMR frequencies of 5.6\,GHz and 4.6\,GHz for the 500 and 200\,nm wide waveguides, respectively. The strain was considered to be uniform in the excitation region and quasi-static. The model therefore neglects effects of phonon propagation and phonon--magnon interactions in the rest of the waveguide. Such a situation can be experimentally realized in good approximation \textit{e.g.} by including a magnetostrictive layer underneath a piezoelectric actuator that is exchange-coupled to an otherwise nonmagnetostrictive waveguide.

\begin{figure*}[p]
  \centering
  \includegraphics[width=170 mm]{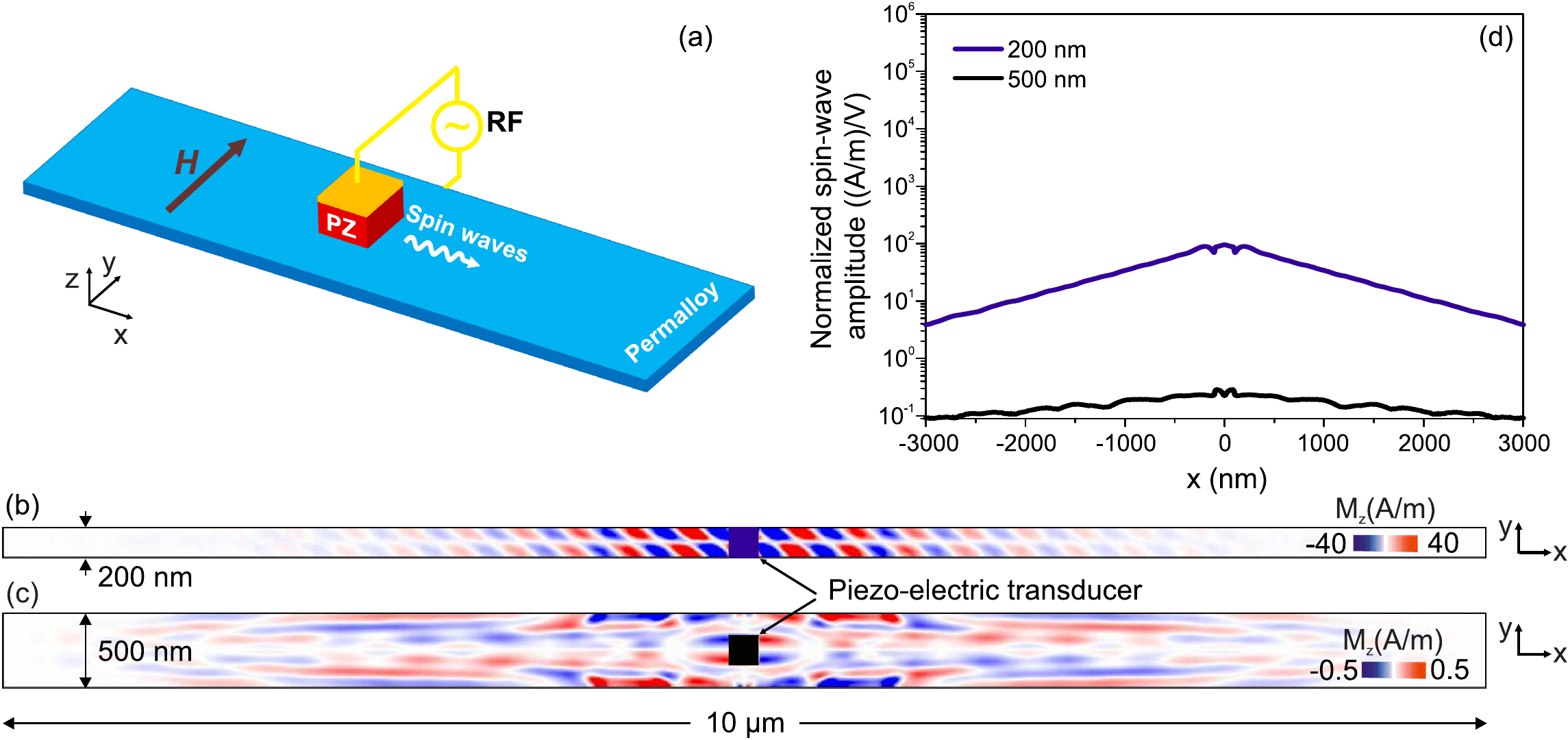}
\caption{\label{fig:Uniaxial}(Color online) Uniaxial normal out-of-plane strain: \textbf{(a)} Device design consisting of a piezoelectric pillar (red) on top of the magnetic waveguide (blue). The piezoelectric pillar is actuated by an rf voltage source via a top electrode. $M_z$ snapshot images of the magnetization oscillation pattern in \textbf{(b)} \SI{200}{\nano\meter} and \textbf{(c)} \SI{500}{\nano\meter} wide waveguides, respectively, after \SI{9}{\nano\second} of excitation by uniaxial out-of-plane strain oscillating at \SI{8}{\giga\hertz}. The static background magnetization was subtracted to enhance the signal. \textbf{(d)} Corresponding amplitudes of the generated spin waves as a function of propagation distance $x$ for both the \SI{200}{\nano\meter} and \SI{500}{\nano\meter} waveguides.}
\end{figure*}

Below, we discuss the magnetoelastic generation of spin waves in three different excitation geometries: (i) uniaxial normal out-of-plane strain, (ii) biaxial normal in-plane strain, and (iii) in-plane shear strain. We first discuss the effect of oscillating uniaxial out-of-plain strain (Fig.~\ref{fig:Uniaxial}) with all components of the strain tensor being zero except $\varepsilon_{zz}$.\cite{Remark} Experimentally, this may be realized by a piezoelectric actuator with a top contact exerting stress on the waveguide underneath [Fig.~\ref{fig:Uniaxial}(a)]. In this geometry, the generated effective anisotropy field is proportional to $m_z = M_z/M_S$ [see Eq.~(\ref{eq:magnetostrictiveeffectivefield})]. 

Figures \ref{fig:Uniaxial}(b) and (c) show the resulting magnetization oscillation pattern ($M_z$ component) in 200\,nm and 500\,nm wide waveguides, respectively, after excitation for 9\,ns with a uniaxial out-of-plane strain oscillating at 8\,GHz. All patterns here and below were obtained in the linear regime, \textit{i.e.} spin wave amplitudes were proportional to the magnitude of the strain. For comparability purposes, all amplitudes were normalized to the exciting voltage. Magnetization pattern are always shown for \SI{1}{\volt} of applied voltage. Detailed quantitative descriptions of the strain tensors used in the simulations can be found in the supplementary information.

The 500\,nm waveguide showed a complex magnetization pattern due to the superposition of multiple spin wave modes. By contrast, the 200\,nm wide waveguide showed a much more uniform wave front. Simulations were also performed with an oscillating external magnetic field, mimicking excitations by the Oersted field of a microwave antenna. In this case, the same mode patterns were observed (data not shown), suggesting that they are inherent to the waveguide rather than to the excitation mechanism. 

In both cases, we thus observe spin wave excitation by the magnetoelastic effect due to uniaxial out-of-plane strain. As shown in Figs.~S1(a) and (b), the magnetization was not uniform due to the nonuniformity of the demagnetizing field. This led to nonzero $m_z$ in the waveguide. However, due to the shape anisotropy of the film, the $m_z$ component was rather small. Therefore, the generated magnetoelastic torques were weak and spin wave amplitudes rather low, as shown in Fig.~\ref{fig:Uniaxial}(d) for both 200\,nm and 500\,nm wide waveguides, respectively. In these graphs, the spin wave amplitude was calculated as the average deviation from the equilibrium magnetization over the cross section of the waveguide and over one excitation period. The spin wave amplitude was found to be much larger in the 200\,nm wide waveguide than in the 500\,nm wide one, where it was essentially negligible. This can be understood by a larger nonuniformity of the demagnetizing field in the narrower waveguide, resulting in increased $m_z$. In all cases, an exponential decay of the spin wave amplitude along the waveguide was observed. According to the analytic calculations of the dispersion relations, the spin waves at \SI{8}{\giga\hertz} have a higher group velocity in 500\,nm wide waveguide when compared to the 200\,nm case, hence the difference in the decay length observed in Fig.~\ref{fig:Uniaxial}(d). 

\begin{figure*}[t]
  \centering
  \includegraphics[width=170 mm]{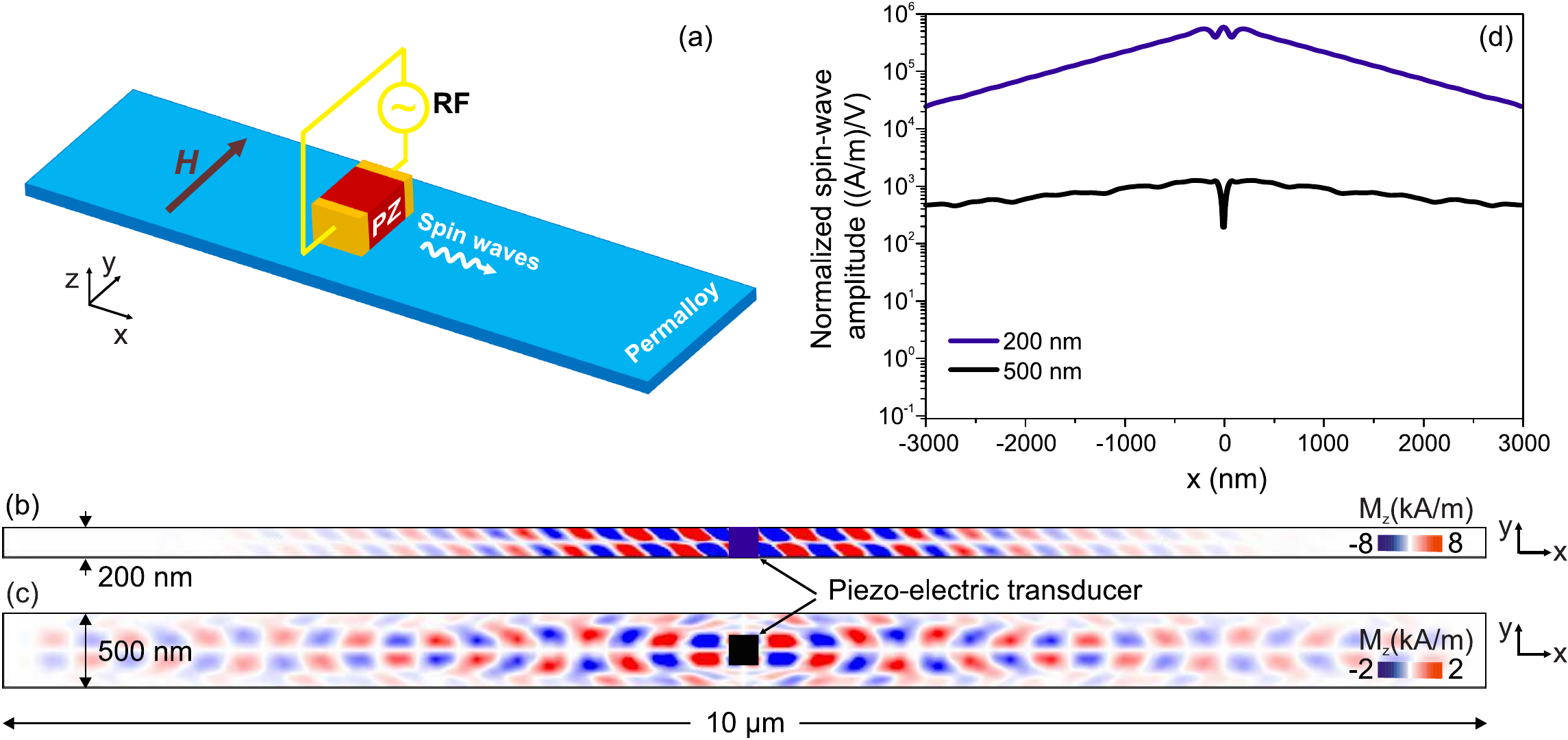}
  \caption{\label{fig:BiaxialEVERYTHING}(Color online) Biaxial normal in-plane strain:\textbf{(a)} Device design consisting of a piezoelectric pillar (red) on top of the magnetic waveguide (blue). The pillar is connected to an rf voltage source via two contacts placed on its sidewalls to generate biaxial in-plane strain. $M_z$ snapshot images of the magnetization oscillation pattern in \textbf{(b)} \SI{200}{\nano\meter} and \textbf{(c)} \SI{500}{\nano\meter} wide waveguides, respectively, after \SI{9}{\nano\second} of excitation by biaxial strain oscillating at \SI{8}{\giga\hertz}. \textbf{(d)} Corresponding amplitudes of the generated spin waves as a function of propagation distance $x$ for both the \SI{200}{\nano\meter} and \SI{500}{\nano\meter} waveguides.}
\end{figure*}

By contrast, the $m_x$ component along the waveguide induced by the effect of the demagnetizing field was much larger than $m_z$ due to the shape anisotropy of the waveguide. As shown in  Eq.~(\ref{eq:magnetostrictiveeffectivefield}), this can be exploited by applying $\varepsilon_{xx}$, which leads to a component of the effective anisotropy field proportional to $m_x$. This situation can be realized by a normal biaxial in-plane strain, \textit{e.g.} experimentally by a piezoelectric actuator with side contacts [Fig.~\ref{fig:BiaxialEVERYTHING}(a)]. In this case, the strain tensor contains nonzero $\varepsilon_{xx}$ and $\varepsilon_{yy}$ components with opposite signs.\cite{Remark}

Figures \ref{fig:BiaxialEVERYTHING}(b) and \ref{fig:BiaxialEVERYTHING}(c) display the distribution of the $M_z$ component of the magnetization in the two waveguides of 200\,nm and 500\,nm width, respectively, after excitation for 9\,ns by oscillating biaxial in-plane strain. The frequency was 8\,GHz, as above. In the 200\,nm wide waveguide, the same spin wave mode as in Fig.~\ref{fig:Uniaxial}(b) was observed, albeit with a much larger amplitude. Moreover, a clear mode pattern became visible in the 500\,nm wide waveguide [Fig.~\ref{fig:BiaxialEVERYTHING}(c)].

Figure~\ref{fig:Uniaxial}(d) shows that the amplitude of spin waves excited by biaxial in-plane strain was about $10^4$ larger than for uniaxial out-of-plane strain. As discussed above, this can be attributed to a much larger $m_x$ component in the waveguide with respect to $m_z$ due to shape anisotropy. In addition, for the narrow waveguide, the demagnetizing field led to a spontaneous symmetry breaking, giving rise to an average $m_x$ in the whole sample, rather than only at the edges (see Fig.~S1 in the supplementary information), further enhancing the spin wave excitation efficiency.

\begin{figure*}[t]
  \centering
  \includegraphics[width=170 mm]{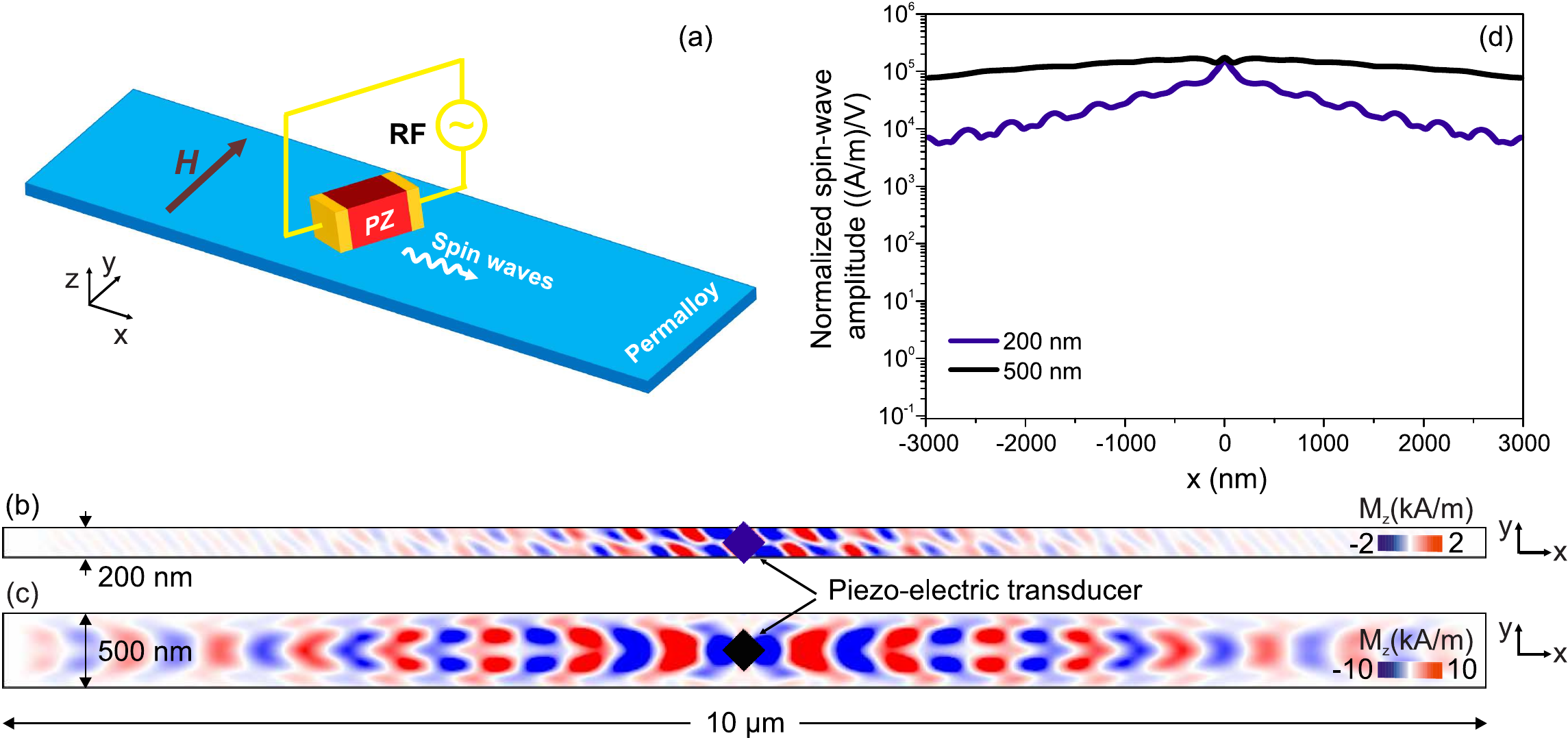}
  \caption{\label{fig:ShearEVERYTHING} (Color online) In-plane shear strain: \textbf{(a)} Device design including a piezoelectric pillar (red) on top of the magnetic waveguide (blue). The pillar is connected to an rf voltage source via two contacts on the sides (yellow) and it is rotated by 45\deg{} to generate shear strain. $M_z$ snapshot images of the magnetization oscillation pattern in \textbf{(b)} \SI{200}{\nano\meter} and \textbf{(c)} \SI{500}{\nano\meter} wide waveguides, respectively, after \SI{9}{\nano\second} of excitation by a shear strain oscillating at \SI{8}{\giga\hertz}. (d) Corresponding amplitudes of the generated spin waves as a function of propagation distance $x$ for both the \SI{200}{\nano\meter} and \SI{500}{\nano\meter} waveguides.}
\end{figure*}

It is instructive to compare the spin wave excitation efficiency in scaled waveguides using biaxial in-plane strain to strain geometries where the torque is nonzero even for uniform magnetization, \textit{e.g.} for in-plane shear strain with nonzero $\varepsilon_{xy}$. In this case, a term proportional to $m_y$ appears in the effective field in Eq.~(\ref{eq:magnetostrictiveeffectivefield}). Such shear strains can be experimentally realized by rotating the piezoelectric actuator used to generate biaxial in-plane strain by 45\deg{} around $z$, as shown in Fig.~\ref{fig:ShearEVERYTHING}(a). A detailed derivation of the used strain tensor and its explicit form can be found in the supplementary information. 

Figures \ref{fig:ShearEVERYTHING}(b) and \ref{fig:ShearEVERYTHING}(c) show snapshot images of $M_z$ for the two waveguides of 200\,nm and 500\,nm width, respectively, after excitation for 9\,ns by in-plane shear strain $\varepsilon_{xy}$ oscillating at 8\,GHz. Amplitudes of the resulting spin waves propagating along the waveguides are shown in Fig.~\ref{fig:ShearEVERYTHING}(d). In contrast to the above cases, the dependence of the spin wave amplitude on the waveguide width was weak since nonuniformities of the demagnetizing field do not play a necessary role in generating torques on the magnetization in this geometry. For a 500 \,nm wide waveguide, shear stress led to about one order of magnitude larger spin wave amplitudes with respect to biaxial in-plane strain. However, for 200\,nm wide waveguides, shear stress was even found to be slightly \textit{less efficient} in exciting spin waves than biaxial in-plane strain, corroborating the strong impact of the demagnetizing field. While a nonzero $m_x$ is strongly beneficial for in biaxial in-plane strain and increases the torque on the magnetization, it actually \textit{decreases} the torque in the case of shear strain, as the main torque component is proportional to $m_y^2 - m_x^2$.\cite{Remark}

In conclusion, we have studied the excitation of spin waves in scaled magnetic waveguides by the magnetoelastic effect using micromagnetic simulations. In the case of a uniform magnetization, normal strains along the principal axes parallel or perpendicular to the magnetization do not lead to torques and therefore cannot excite spin waves. In scaled waveguides, the effects of nonuniform demagnetizing fields lead however to nonzero torques and spin wave generation even for normal strain along principal axes. Biaxial in-plane strain was found to be about four orders of magnitude more efficient than uniaxial out-of-plane strain. In 200\,nm wide waveguides, biaxial in-plane stress was found to be even more efficient than shear stress that leads to nonzero torques even in uniformly magnetized waveguides. This indicates that magnetoelectric spin wave transducers using biaxial in-plane strain may be highly efficient to excite spin waves in scaled waveguides in a Damon-Eshbach geometry without the need to generate shear strains.

See the supplementary material for an expression of the magnetoelastic torque, detailed quantitative descriptions of the strain tensors used in the simulations, including their derivations, as well as the components of the magnetization in 200 and 500\,nm wide waveguides.

\clearpage

\clearpage

\clearpage

\end{document}